\documentclass[aps,prl,twocolumn,groupedaddress,showpacs,floatfix]{revtex4}
\def\bc{\begin{center}}
\def\ec{\end{center}}
\def\be{\begin{equation}}
\def\ee{\end{equation}}
\renewcommand{\vec}[1]{\mbox{\boldmath$#1$}}
\usepackage{longtable}
\usepackage{epsfig}
\begin{document}

\title{Microscopic verification of topological electron-vortex binding in the 
lowest-Landau-level crystal state}
\author{Chia-Chen Chang, Gun Sang Jeon, and Jainendra K. Jain}
\affiliation{Department of Physics, 104 Davey Laboratory, The Pennsylvania State University,
Pennsylvania 16802}
\date{\today}

\begin{abstract}
When two-dimensional electrons are subjected to a very strong
magnetic field, they are believed to form a triangular crystal.
By a ditect comparison with the {\em exact} wave function, we 
demonstrate that this crystal is not a 
simple Hartree-Fock crystal of electrons but an inherently quantum 
mechanical crystal characterized by a non-perturbative binding of 
quantized vortices to electrons.  It is suggested that 
this has qualitative consequences for experiment.
\end{abstract}

\maketitle

The quantum mechanical behavior of solids has been subject 
of much investigation in the context of the quantum solids 
of $^3$He and $^4$He~\cite{Cross}.  There has been a revival of interest 
in this topic due to the recently reported ``supersolid" phase of 
$^4$He~\cite{Chan}, resulting from a significant 
overlap between the wave functions of neighboring atoms~\cite{Leggett}.
This paper demonstrates that the lowest Landau level (LL) crystal of 
electrons provides another example of an inherently quantum mechanical 
crystal.

Much work has been done on the lowest LL crystal for over two 
decades.  When a two-dimensional electron system is exposed to a magnetic 
field ($B$), the kinetic energy is quantized into LL's.  
The number of occupied LL's is called the 
filling factor, $\nu=\rho hc/eB$, $\rho$ being the two-dimensional 
electron density.  At sufficiently strong magnetic fields, 
when all electrons fall into the lowest LL ($\nu<1$), 
the kinetic energy is no longer relevant, and the nature of the 
state is determined solely by the Coulomb interaction.
Following Wigner~\cite{Wigner}, the dominance of the interaction 
energy can be expected to produce an electron crystal.  
For a range of filling factors the system condenses into a quantum 
liquid, characterized by dissipationless transport and precisely quantized plateaus of 
Hall resistance~\cite{Tsui}.  There are strong indications, however, that a 
crystal occurs at sufficiently low $\nu$,~\cite{Shayegan} and its properties 
have been probed experimentally by transport~\cite{Jiang,Goldman,Li,Santos,Pan} and by  
electromagnetic waves~\cite{Andrei,Buhmann,Williams,Kukushkin,Ye,Chen}, as well as  
theoretically~\cite{Lozovik,Fukuyama,Maki,Levesque,Lam,Chui,Zhu,Price,Zheng,Fertig,Murthy,Haldane,Peterson}.

Certain microscopic wave functions~\cite{Laughlin,Jain} are known to provide 
a good account of the  fractional-quantum-Hall-effect (FQHE) liquid, 
as ascertained from comparisons with exact solutions known for systems containing 
a finite number of electrons, but become progressively worse with decreasing 
$\nu$.  A Hartree-Fock wave function describing an ``electron 
crystal" (EC) provides a better approximation 
for the ground state at low $\nu$.  However, the 
Hartree-Fock crystal is not particularly good either, which has 
raised questions regarding the true nature of the crystal.  
A very interesting proposal suggests that the physics 
underlying the FQHE liquid is also operative in  
the crystal phase~\cite{Fertig,Murthy}.  Yi and Fertig~\cite{Fertig} 
have shown that a variational wave function in which vortices are bound 
to electrons has lower energy than the earlier Lam-Girvin~\cite{Lam}
wave function in the filling factor range $0.1<\nu<0.2$.
Narevich, Murthy, and Fertig~\cite{Murthy} have used a  
Hamiltonian formulation of composite fermions to estimate gaps 
and shear modulus on either side of the $\nu=1/5$ quantum Hall state.

The notion of binding of quantized vortices to electrons in the lowest 
LL crystal, if confirmed, would indicate the formation of a quantum crystal, 
given that vortices are inherently quantum mechanical objects.
While the consequences of the quantum mechanical nature of such a  
crystal ought to be evaluated and tested by experiment, rigorous and 
unbiased theoretical tests of electron-vortex binding are possible 
because of the fortunate feature that the {\em exact} ground state wave 
function can be obtained, for finite systems, by a brute force numerical 
diagonalization for a wide range of $\nu$ in the crystal phase.
The principal result of this work is to show that a wave function for the 
composite-fermion (CF) crystal, the composite fermion being the 
bound state of an electron and an even number of vortices,
is extremely accurate at low $\nu$ -- 
more accurate than the accepted FQHE wave functions for 
the liquid phase -- thus establishing that 
the real crystal indeed has vortices bound to electrons.  
We also determine the parameter range where the CF crystal occurs.
One might have expected 
the physics of the liquid to carry over into the crystal phase in the proximity of 
the phase boundary separating the liquid and the crystal, but 
our calculations indicate that the CF crystal is realized even deep inside 
the crystal phase, down to the lowest $\nu$ considered below.

The wave packet for an electron in the lowest LL 
localized at $\vec{R}=(X,Y)$ is given by~\cite{Maki} 
\be
\phi_{\vec{R}}(\vec{r})=\frac{1}{\sqrt{2\pi}} \exp\left(-\frac{1}{4}(\vec{r}-\vec{R})^2+\frac{i}{2}
(xY-yX)\right)
\ee
where the magnetic length, $l_0=\sqrt{\hbar c/eB}$, has been taken as the unit of length.
The wave function for the electron crystal (EC) is constructed by 
placing electrons on a triangular lattice $\vec{R}_j$, the lowest 
energy solution for the classical problem, and then antisymmetrizing
the product:~\cite{Maki} 
\be
\Psi^{EC}=\frac{1}{\sqrt{N!}}\sum_P \epsilon_P \prod_{j=1}^N \phi_{\vec{R}_j}(\vec{r}_{Pj})
\ee
where the sum is over all permutations $P$ and $\epsilon_P$ is $+1$ for even 
permutations and $-1$ for odd permutations.
With the lattice constant $a=(4\pi/\sqrt{3}\nu)^{1/2}l_0$, 
the overlap integral between nearest neighbor electron 
wave functions~\cite{Maki}, $\exp(-a^2/2l_0^2)=\exp(-3.627/\nu)$, 
decays rapidly with decreasing $\nu$.  
We will work with the symmetric gauge, $\vec{A}=(B/2)(-y,x,0)$, for which 
the total angular momentum $L$ is a good quantum number.
Because the wave function $\Psi^{EC}$ is not an eigenstate of angular momentum,
we follow the method of Yannouleas and Landman~\cite{Landman} to project 
it onto a definite $L$, denoting the resulting wave function  $\Psi^{EC}_{L}$.
Such projection amounts to creating a rotating crystal, implying that 
the crystalline structure is not apparent in the density but
in the pair correlation function.  The explicit expression 
for $\Psi^{EC}_{L}$ is given in Ref.~\onlinecite{Landman}.

Following the standard procedure of 
the composite fermion (CF) theory~\cite{Jain,Kamilla}, we construct the following 
wave function:
\be
\Psi^{^{2p}CFC}_L=\prod_{j<k}(z_j-z_k)^{2p}\Psi^{EC}_{L^*}
\label{CFC},
\ee 
\be
L^*=L-pN(N-1).
\ee
It is interpreted as a CF crystal (CFC),
because the Jastrow factor $\prod_{j<k}(z_j-z_k)^{2p}$ binds $2p$ 
quantized vortices to each electron in $\Psi^{EC}$ to convert it into a 
composite fermion; the composite fermions of different flavors are denoted by 
$^{2p}$CF, and their crystals by $^{2p}$CFC.  We next proceed to compare 
$\Psi^{^{2p}CFC}$ with exact wave functions.  The latter  
can be obtained (using the Lanczos method) for up to $N=7$ particles
in the low-$\nu$ region of interest.
We will present below detailed results for $N=6$; 
the study of $N=5$ and $N=7$ particles is consistent with 
our conclusions below.  The filling factor of the finite system will 
be defined by the expression $\nu=N(N-1)/2L$, which gives the 
correct value of $\nu$ for a uniform density state 
in the thermodynamic limit.  For $N=6$, the lowest 
energy classical configuration has one particle at the center, with 
the remaining five forming a ring around it~\cite{Peeters}.
The wave functions $\Psi^{^{2p}CFC}$ for $2p\neq 0$ have rather complicated 
correlations built into them, but the interaction energy per particle,
\be
V=\frac{1}{N}\frac{\langle\Psi^{^{2p}CFC}_L| \sum_{j<k} 
\frac{e^2}{\epsilon r_{jk}}|\Psi^{^{2p}CFC}_L \rangle}
{\langle\Psi^{^{2p}CFC}_L|\Psi^{^{2p}CFC}_L \rangle}  \;, 
\ee
can be evaluated by the Metropolis Monte Carlo method at least for many large values 
of $2p$ (the computation time increases rapidly as $2p$ is reduced). 
The total energy also has contributions from  
electron-background and background-background interactions, but these terms 
are the same for different crystal wave functions for a given $L$, so are not  
relevant for comparisons.

\begin{figure}
\centerline{\epsfig{file=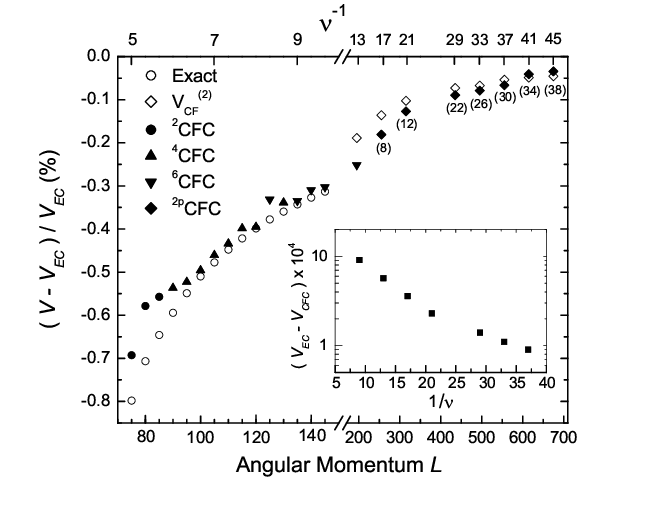,width=8cm}}
\caption{
\label{fig1}
The correlation energy of the optimal CF crystal, i.e., the 
\% deviation of its Coulomb energy from the Coulomb energy of the 
uncorrelated electron crystal, for $N=6$ particles.
The superscript $2p$ on $^{2p}$CFC indicates the
vortex quantum number of composite fermions.  The energy of the electron crystal
for $L>400$ is taken from Yannouleas and Landman~\protect\cite{Landman}.
The deviation of the exact energy from the electron crystal energy is also 
shown for $L\leq 145$; for larger angular momenta, where the exact energy is 
not available, we show an accurate approximation, $V_{CF}^{(2)}$
(explained in the text), as an independent reference.  For $2p>6$, the
number of vortices carried by composite fermions 
is shown in brackets near the diamond.  The energy difference
per particle between the electron and the CF crystals is given in the inset,
quoted in units of  $e^2/\epsilon l_0$,
where $l_0$ is the magnetic length and $\epsilon$ the dielectric constant of
the host semiconductor.
}
\end{figure}

Fig.~\ref{fig1} shows the correlation energy of the optimal CF crystal, 
defined as the deviation of its energy from that of the uncorrelated electron crystal.
(The computationally accessible range of $2p$ allows us to determine 
the minimum CFC energy.  The only exceptions are the largest two values of $L$, 
where we show the energy at the smallest $2p$ studied; the minimum CF energy may 
be still lower here.)  The exact correlation energy is also shown for $L\leq 145$.
For $L>145$,  the dimension of the Fock space ($D$) is too large for an exact treatment.
As an independent reference point, we 
obtain an accurate approximation to the exact energy by the method of ``CF diagonalization,"
wherein the Coulomb Hamiltonian is diagonalized in a correlated CF basis, the dimension of
which is much smaller than the dimension of the full basis needed
for the exact state; gradually increasing the
basis size gives an increasingly better approximation~\cite{Jeon}.
Fig.~\ref{fig1} quotes $V^{(2)}_{CF}$ (using the notation in Ref.~\onlinecite{Jeon}),
obtained with a correlated CF basis of dimension 150.   $V^{(2)}_{CF}$ has been 
shown to be very precise~\cite{Jeon}: for the six particle system 
it is within 0.02\% of the exact energy for $L\leq 145$ and we expect 
similar level of accuracy for higher $L$ as well.

The minimum energy for all $L$ is obtained at a non-zero 
value of $2p$, which establishes that the CF crystal  
provides a better variational state than the electron crystal.  
Most significantly, the CF crystal is essentially the 
exact solution for $\nu \leq 1/7$ ($L\geq 105$).  
For $100 < L < 145$, the energy of the optimal  
CF crystal is approximately within $0.02$\% of the exact energy.  Tables~\ref{Otable} 
and \ref{Etable} show how Laughlin's liquid wave function, $\Psi^{EC}$, and $\Psi^{CFC}$ 
compare with the exact wave function for $\nu=1/5$, $1/7$ and $1/9$. 
As indicated earlier, the liquid wave function worsens and $\Psi^{EC}$ improves with 
decreasing $\nu$, but neither is very good at small fillings.
In contrast, $\Psi^{CFC}$ is surprisingly close to the exact state.
Its overlap with the exact wave function is $\sim$99\% for $\nu=1/7$ and $1/9$, while 
its energy deviates from the exact energy 
by 0.016\% and 0.006\%, respectively.  It is worth noting: (i)  
The exact state is a linear combination of a large number of Slater determinant
basis functions (see Table~\ref{Otable}), involving $D-1$ parameters,
and yet, a single CFC wave function captures its physics almost exactly.
(ii) The CFC wave function for $\nu\leq 1/7$ is more accurate than 
Laughlin's wave function at $\nu=1/3$, whose energy for $N=6$ (in the disk 
geometry) is off by 0.15\% and whose overlap with the exact state is 0.964, 
in spite of the fact that the dimension of the Fock space at $\nu=1/3$ is much 
smaller ($D=1206$).  For larger $L$, the energy of the CFC is 
lower than  $V^{(2)}_{CF}$, with the possible exception of the last two points, 
where we may not have the optimal CFC.

\begin{table}
\caption{The last three columns give the 
overlaps of CF crystal (CFC), electron crystal (EC), and Laughlin's
wave function with the exact ground state wave function at several filling factors $\nu$.
The overlap is defined as $|\langle
\Psi^{trial}|\Psi^{exact}\rangle|^2 /
\langle \Psi^{trial}|\Psi^{trial}\rangle \langle \Psi^{exact}|\Psi^{exact}\rangle$.
The second column gives $D$, the dimension of the basis space for $N=6$ electrons, 
and $L$ is the total angular momentum of the state.
\label{Otable}
}
\begin{tabular}{|c | c | c | c | c|}\hline
$ \nu\;\;\; (L) $  & $D$ & CFC & EC & Laughlin \\ \hline \hline
1/5 $\phantom{0}$(75) & $\phantom{0}$19858 & 0.891 & 0.645 & 0.701    \\ \hline 
            1/7 (105)  & 117788 & 0.994 & 0.723 & 0.504    \\ \hline 
            1/9 (135) & 436140 & 0.988 & 0.740 & 0.442    \\ \hline 
\end{tabular}
\end{table}

\begin{figure}
\centerline{\epsfig{file=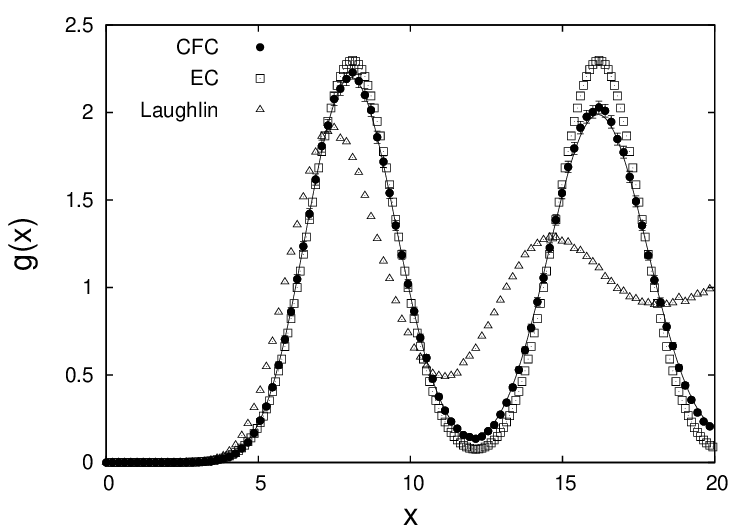,width=8cm}}
\caption{
\label{fig:pair}
The pair correlation functions for the CF crystal (solid circles), the electron crystal
(empty squares), and Laughlin's wave function (empty triangles)
on a circle of radius $R=6.445l_0$ for six particles at $\nu=1/7$.
The solid line shows the exact pair correlation function.
}
\end{figure}

Because every particle sees quantized vortices on every other particle, the formation 
of composite fermions implies a long range quantum coherence in the 
crystal phase.  To get a feel for how the binding of vortices to 
electrons affects the inter-particle correlations, we show in Fig.~\ref{fig:pair} 
the pair correlation function $g(x)$
for several candidate wave functions as well as the exact ground state 
for $\nu=1/7$; $g(x)$ is the probability 
of finding a pair of particles at an arc distance $x$ on a circle of radius $R$.
($R$ is chosen to match the distance of a particle in the parent classical 
crystal from the center of the disk.) The result shows that the 
crystalline correlations are slightly weakened by the formation of 
composite fermions.  It is perhaps counter-intuitive that such an effect 
should lead to a lower energy even at very low fillings.

\begin{table}
\caption{\label{Etable}
Interaction energies per particle for the exact ground state, the CF crystal (CFC), 
the electron crystal (EC), and Laughlin's wave function for six particles at 
several filling factors.  The uncertainty in the last digit from Monte Carlo sampling 
is given in parentheses.
}
\vspace*{0.2in}
\begin{tabular}{| c|  c|  c | c | c|}\hline
$ \nu\;\;\; (L) $           & exact  &  CFC & EC &  Laughlin \\ \hline \hline
1/5 $\phantom{0}$(75) & 2.2019 & 2.2042(5) & 2.2196 & 2.2093(2) \\ \hline 
            1/7 (105) & 1.8533 & 1.8536(2) & 1.8622 & 1.8617(2) \\ \hline 
            1/9 (135) &1.6305 & 1.6306(1) & 1.6361 & 1.6388(1) \\ \hline 
\end{tabular}
\end{table}

Of interest is the nature of the thermodynamic state, 
obtained in the limit $N\rightarrow \infty$ at a fixed filling factor. 
Finite size studies do not necessarily provide a reliable account of  
the thermodynamic state.  For example, for $N=6$ 
the CFC gives a better description of the $\nu=1/5$ ground state 
than Laughlin's liquid wave function, even though the thermodynamic 
state here is known to be a liquid~\cite{Jiang,Goldman}.  
However, when an extremely precise and unambiguous description of 
the finite $N$ state is obtained, as is the case at $\nu\leq 1/7$,  
we consider that to be a strong indication for the 
nature of the state in the thermodynamic limit. 
In any case, even though our finite $N$ study cannot give the precise $\nu$ value  
where a transition from liquid to crystal takes place, it does make a compelling 
case that whenever the thermodynamic state is a crystal, it is a crystal of 
composite fermions, even in regions of the phase diagram far from the CF liquid.

The quantum character of the crystal is not fragile, and ought to be observable at 
presently attainable temperatures, even at very small $\nu$. 
The energy difference per particle, $V^{CFC}-V^{EC}$, shown in the 
inset of Fig.~\ref{fig1}, gives a crude estimate for the temperature below 
which the quantum nature of the crystal should be robust to thermal 
fluctuations.  The relevant temperatures appear to be 
well within the present experimental reach -- for example, for 
parameters appropriate for GaAs, the quantum 
crystal regime is estimated to be below $\approx$ 25 mK (at $B=25$T) 
even at $\nu=1/33$.  From the $N$ dependence,  
we have estimated that the energy difference shown in the inset 
underestimates the thermodynamic energy difference by approximately 
a factor of two.  It is interesting to note that even as the 
energy difference between the CF and the electron
crystals decreases as $\nu\rightarrow 0$, $2p$ continues to rise.  
Thus, CF flavors of up to very high $2p$ are predicted to occur in the 
crystal state.  In the liquid phase, $^2$CFs and $^4$CFs have 
definitely been observed, and there is evidence also for $^6$CFs 
and $^8$CFs at relatively high temperatures.~\cite{Pan}  

Unlike in bosonic quantum crystals, 
the overlap between (uncorrelated) electron wave packets at 
neighboring sites is negligible in the filling factor region of interest 
(the overlap integral is $10^{-15}$ for $\nu=1/9$).  The quantum nature 
of the CF crystal owes its origin to the long range Coulomb interaction.

Given that the CF liquid behaves qualitatively differently from 
an electron liquid, one may ask in what ways the properties of the 
CF crystal are distinct from those of an electron crystal.
We mention here a few examples where the CFC can provide natural 
explanations for certain 
experimental facts, although further work will be needed to make the 
connection with experiment more direct and to clarify other possible  
implications.  The issue is obviously  
relevant to experiments that exhibit transitions between 
the liquid and crystal phases.  Re-entrant transitions between 
the FQHE liquid and an insulating state, thought to be a pinned 
crystal, have been observed~\cite{Jiang,Goldman} in going from  
$\nu=2/5$ to $\nu=1/5$.  In the filling factor range $1/5>\nu>1/9$,
the low temperature insulating state melts into a CF liquid upon raising 
temperature~\cite{Pan}, as indicated by the appearance of FQHE like
structure.  These observations become less baffling knowing that the 
crystal is itself made of composite fermions rather than electrons,
thus requiring a less drastic reorganization of the state at the transition.   
Another result, perhaps puzzling for an electron crystal, is that the Hall 
resistance of the pinned crystal is close to the value it would have for 
a liquid~\cite{Goldman2}.  If the current is carried by composite 
fermions instead, then the Hall voltage 
induced by the accompanying vortex current (the vortices effectively behave
as magnetic flux quanta~\cite{Jain,Lopez}), through an effective 
Faraday effect, is roughly consistent with the observation.  
(Zheng and Fertig~\cite{Zheng} considered a similar mechanism for transport 
by correlated interstitial defects.)  The unexpectedly 
small activation energy in the crystalline state, compared to theoretical 
predictions based on an electron crystal, as well as its non-monotonic 
filling factor dependence has also been  
rationalized in terms of a CF crystal~\cite{Fertig,Murthy}.

Partial support of this research by the National Science Foundation 
under grant no. DMR-0240458 is gratefully acknowledged.
We thank H.A.  Fertig and G. Murthy for useful discussions and their 
comments on the manuscript.


\begin{thebibliography}{99}

\bibitem{Cross} M. C. Cross and D. S. Fisher
Rev. Mod. Phys. {\bf 57}, 881 (1985); D.M. Ceperley, Rev. Mod. Phys. {\bf 67}, 
279 (1995).

\bibitem{Chan} 
E. Kim and M.H.W. Chan, Nature {\bf 427}, 225 (2004).

\bibitem{Leggett} 
A.J. Leggett, Phys. Rev. Lett.  {\bf 25}, 1543 (1970).

\bibitem{Wigner} 
E.P. Wigner, Phys. Rev. {\bf 46}, 1002 (1934). 

\bibitem{Tsui} 
D.C. Tsui, H.L. Stormer, and A.C. Gossard, 
Phys. Rev. Lett. {\bf 48}, 1559 (1982).

\bibitem{Shayegan} For a review, 
see articles by M. Shayegan and H.A. Fertig in {\em Perspectives in Quantum Hall Effects}, 
S. Das Sarma, A. Pinczuk, Eds.  (Wiley, New York, 1997).

\bibitem{Jiang} 
H.W. Jiang {\it et al.}, 
Phys. Rev. Lett. {\bf 65}, 633 (1990). 
 
\bibitem{Goldman} 
V.J. Goldman {\it et al.},
{\it Phys. Rev. Lett.} {\bf 65}, 2189 (1990). 

\bibitem{Li} Y. P. Li {\em et al.}, 
Phys. Rev. Lett. 67, 1630 (1991).

\bibitem{Santos} M. B. Santos {\em et al.}, 
Phys. Rev. Lett. 68, 1188 (1992).

\bibitem{Pan} 
W. Pan {\it et al.}, Phys. Rev. Lett. {\bf 88}, 176802 (2002). 

\bibitem{Andrei} E.Y. Andrei {\em et al.}, Phys. Rev. Lett. {\bf 60}, 2765 (1988).

\bibitem{Buhmann} H. Buhmann {\em et al.}, Phys. Rev. Lett. {\bf 66}, 926 (1991).

\bibitem{Williams}
F.I.B. Williams {\em et al.}, Phys. Rev. Lett. {\bf 66}, 3285 (1991).

\bibitem{Kukushkin} I. V. Kukushkin {\em et al.}, 
Phys. Rev. Lett. {\bf 72}, 3594 (1994).

\bibitem{Ye} P. D. Ye {\em et al.}, 
Phys. Rev. Lett. {\bf 89}, 176802 (2002).

\bibitem{Chen} Y. Chen {\em et al.} Phys. Rev. Lett. {\bf 91}, 016801 (2003).

\bibitem{Lozovik} Y.E. Lozovik and V.I. Yudson, JETP Lett. {\bf 22}, 11 (1975).

\bibitem{Fukuyama} H. Fukuyama, P.M. Platzman, and P.W. Anderson,
Phys. Rev. {\bf B 19}, 5211 (1979).

\bibitem{Maki} 
K. Maki and X. Zotos, Phys. Rev. B {\bf 28}, 4349 (1983).

\bibitem{Levesque} D. Levesque, J.J. Weis, and A.H. MacDonald, Phys. Rev. 
B {\bf 30}, 1056 (1984).

\bibitem{Lam} 
P.K. Lam and S.M. Girvin, Phys. Rev. B {\bf 30}, 473 (1984).

\bibitem{Chui} K. Esfarjani and S.T. Chui, Phys, Rev.  B {\bf{42}}, 10758
(1990).

\bibitem{Zhu}
X. Zhu and S. G. Louie, Phys. Rev. Lett. {\bf 70}, 335 (1993).

\bibitem{Price} 
R. Price, P. M. Platzman, and S. He, Phys. Rev. Lett. {\bf 70}, 339 (1993);
R. Price {\em et al.}, Phys. Rev. B {\bf 51}, 2017 (1996).

\bibitem{Zheng} 
L. Zheng and H.A. Fertig, Phys. Rev. Lett. {\bf 73}, 878 (1994).

\bibitem{Fertig}  
H. Yi and H.A. Fertig, Phys. Rev. B {\bf 58}, 4019 (1998). 

\bibitem{Murthy} 
R. Narevich, G. Murthy, and H.A. Fertig, 
Phys. Rev. B {\bf 64}, 245326 (2001). 

\bibitem{Haldane} K. Yang, F.D.M. Haldane, and E.H. Rezayi, Phys. Rev. B
{\bf{64}}, 081301(2001).

\bibitem{Peterson} 
S.S. Mandal, M.R. Peterson, and J.K. Jain,
Phys. Rev. Lett. {\bf 90}, 106403 (2003). 

\bibitem{Laughlin} R.B. Laughlin, Phys. Rev. Lett. {\bf 50}, 1395 (1983).

\bibitem{Jain}  
J.K. Jain, Phys. Rev. Lett. {\bf 63}, 199 (1989); Phys. Rev. B {\bf 40},
8079 (1989).


\bibitem{Landman} 
C. Yannouleas and U. Landman, Phys. Rev. B {\bf 68}, 035326 (2003);
{\em ibid.} {\bf 69}, 113306 (2004).

\bibitem{Kamilla} 
J.K. Jain and R.K. Kamilla, Int. J. Mod. Phys. B {\bf 11}, 2621 (1997);
Phys. Rev. B {\bf 55}, R4895 (1997). 


\bibitem{Peeters}
V.M. Bedanov and F.M. Peeters, Phys. Rev. B {\bf 49}, 2667 (1994).

\bibitem{Jeon} 
G.S. Jeon, C.-C. Chang, and J.K. Jain, Phys. Rev. B {\bf 69}, 
241304(R) (2004).


\bibitem{Goldman2} 
V.J. Goldman {\em et al.}, Phys. Rev. Lett. {\bf 70}, 647 (1993).

\bibitem{Lopez} A. Lopez and E. Fradkin, Phys. Rev. B {\bf 44}, 5246 (1991); 
B.I. Halperin, P.A. Lee, and N. Read, Phys. Rev. B {\bf 47}, 7312 (1993);
A.S. Goldhaber and J.K. Jain, Phys. Lett. A {\bf 199}, 267 (1995).


\end{thebibliography}
\end{document}